\documentclass[10pt,conference]{IEEEtran}
\IEEEoverridecommandlockouts
\usepackage{cite}
\usepackage{amsmath,amssymb,amsfonts}
\usepackage{algorithmic,float}
\usepackage{algorithm}
\usepackage{graphicx}
\usepackage{textcomp}
\usepackage{xcolor}
\def\BibTeX{{\rm B\kern-.05em{\sc i\kern-.025em b}\kern-.08em
    T\kern-.1667em\lower.7ex\hbox{E}\kern-.125emX}}

\IEEEoverridecommandlockouts\IEEEpubid{\makebox[\columnwidth]{ 978-1-6654-3540-6/22/\$31.00 ~\copyright~2022 IEEE \hfill} \hspace{\columnsep}\makebox[\columnwidth]{ }}

\DeclareRobustCommand*{\IEEEauthorrefmark}[1]{%
  \raisebox{0pt}[0pt][0pt]{\textsuperscript{\footnotesize #1}}%
}

\begin{document}

\title{Deadline-constrained Multi-resource Task Mapping and Allocation for Edge-Cloud Systems
\thanks{This work was supported in part by the MoE Tier-2 grant MOE-T2EP20221-0006.}
}

\author{\IEEEauthorblockN{Chuanchao Gao\IEEEauthorrefmark{1,2}
, Aryaman Shaan\IEEEauthorrefmark{1}, Arvind Easwaran\IEEEauthorrefmark{1,2}}
\IEEEauthorblockA{\IEEEauthorrefmark{1}\textit{School of Computer Science and Engineering}}
\IEEEauthorblockA{\IEEEauthorrefmark{2}\textit{Energy Research Institute @ NTU, Interdisciplinary Graduate Programme}}
\IEEEauthorblockA{
\textit{Nanyang Technological University}\\
Singapore \\
gaoc0008@e.ntu.edu.sg, shaa0002@e.ntu.edu.sg, arvinde@ntu.edu.sg}
}

\maketitle

\begin{abstract}
In an edge-cloud system, mobile devices can offload their computation intensive tasks to an edge or cloud server to guarantee the quality of service or satisfy task deadline requirements. However, it is challenging to determine where tasks should be offloaded and processed, and how much network and computation resources should be allocated to them, such that a system with limited resources can obtain a maximum profit while meeting the deadlines. A key challenge in this problem is that the network and computation resources could be allocated on different servers, since the server to which a task is offloaded (e.g., a server with an access point) may be different from the server on which the task is eventually processed. To address this challenge, we first formulate the task mapping and resource allocation problem as a non-convex Mixed-Integer Nonlinear Programming (MINLP) problem, known as NP-hard. We then propose a zero-slack based greedy algorithm (ZSG) and a linear discretization method (LDM) to solve this MINLP problem. Experiment results with various synthetic tasksets show that ZSG has an average of $2.98\%$ worse performance than LDM with a minimum unit of 5 but has an average of $6.88\%$ better performance than LDM with a minimum unit of 15.
\end{abstract}


\section{Introduction}
Compute intensive tasks are rapidly emerging with the development of Internet of Things and Artificial Intelligence technologies, and this coupled with the deadline requirements of time-critical tasks, introduce a big challenge for systems. For example, in autonomous driving applications, tasks such as object detection and localization fall into this category, and the vehicles (\emph{end devices}) are required to service these tasks while meeting their deadlines. The multi-layer edge-cloud system is often deployed to enhance the end devices' capability of handling such tasks, and such capability will be increased further with the advent of wireless technologies that are capable in handling strict deadlines such as 5G-URLLC~\cite{li20185g}.

In a multi-layer edge-cloud system, tasks can be offloaded from end devices to \textit{access points}, and then forwarded to \textit{servers} for timely processing.
Servers that are located far away from the end devices, which results in significant data transmission latency, are called {cloud servers}.
Servers that are deployed collectively with access points to provide a quick response to end devices are called {edge servers}.
The computation capacity of a cloud server is usually much greater than that of an edge server.
If the tasks received by access points have significant demand for computation resource, the collectively deployed edge servers may not have enough computation resource to finish these tasks by their deadlines. In such a case, the access points can forward the tasks with high computation resource demand to cloud servers for processing.

End devices communicate with access points through a wireless network, and access points communicate with servers through a wired backhaul network that has a much larger bandwidth capacity than the wireless network. Due to the limited wireless bandwidth and computation resource, it is challenging to determine where the tasks should be offloaded and processed (\emph{task mapping problem}), and how much bandwidth and computation resource should be allocated to them (\emph{resource allocation problem}), to maximize system profit while meeting task deadlines.
This problem is further compounded by the fact that the access point that a task is offloaded to and the server that it is eventually processed on may be deployed at different locations.
An access point will allocate bandwidth to tasks that are offloaded to it, and a server will allocate computation resource to tasks that are processed on it.

In this paper, we formulate the above problem as a nonconvex MINLP. Nonconvexity arises due to the deadline constraint, since the allocated wireless bandwidth (respectively computation) has an inverse relation to the time taken for offloading (respectively processing). In our model, end devices can offload tasks to one of several nearby access points using the allocated wireless bandwidth, and each task can be processed on any reachable server using the allocated computation resource. Besides, there is an additional transmission delay incurred by the task if the offloading access point and the processing server are deployed at different locations.
This introduces further challenges because the end-to-end deadline now depends on three factors: 1) wireless bandwidth allocated by the offloading access point, 2) transmission delay between the offloading access point and the processing server, and 3) computation resource allocated by the processing server. A task mapping and resource allocation is deemed \emph{feasible} in this model if the task can be completed by its deadline with the allocated bandwidth and computation resource, inclusive of any transmission delays.

This paper aims to maximize the total system profit, where each task can contribute to this profit only if its allocation is feasible. From the literature on knapsack problems~\cite{chekuri2005polynomial}, the above problem can be categorized as a Generalized Assignment Problem (GAP) with fixed profit and bin-specific sizes for each item, assuming either the wireless bandwidth or the computation resource allocation is fixed. The intuition is that in order to meet task deadlines, the allocation of bandwidth and computation resource depends on the access point and the server to which the task is mapped and the transmission delay between them. Note that GAP is known to be NP-Hard and more specifically APX-Hard~\cite{chekuri2005polynomial}. The contributions of this paper are as follows.
\begin{itemize}
	\item We formulate the deadline-constrained task mapping and resource allocation problem with communication and computation contention as a nonconvex MINLP.
	\item We propose a zero-slack based greedy heuristic algorithm (ZSG) for the above problem, and the resources allocated to all provisioned tasks are just enough for these tasks to be completed exactly at their respective deadlines. We also propose a linear discretization method (LDM) to reformulate the nonconvex MINLP problem into an Integer Linear Programming problem, assuming that the bandwidth and computation resources can only be allocated in discrete units.
	\item We conduct experiments with synthetically generated tasksets to evaluate the performance of the two proposed methods. Results show that ZSG can obtain $2.98\%$ less profit than LDM with a minimum unit of 5 and $6.88\%$ more profit than LDM with a minimum unit of 15.
	Further, the performance of LDM critically depends on the size of the minimum discrete unit that can be allocated; the achieved profit drops by $9.87\%$ on an average when the minimum unit is increased from $5$ to $15$.
\end{itemize}

\noindent \textbf{Related Work.} Few studies have considered this deadline-constrained problem with both computation and communication contention, aimed at minimizing either the total system cost or energy consumption~\cite{vu2021optimal,li2019cooperative}. However, these studies assumed that tasks could be directly offloaded to the servers where they are processed, and hence the multi-resource contention is modeled on the same server for each task. Other studies have considered similar deadline-constrained problems with either a fixed task to server mapping~\cite{millnert2018achieving} or a fixed resource allocation for each task~\cite{cziva2018dynamic}. Finally, a task mapping and computation resource allocation problem with deadlines has also been considered~\cite{yang2019efficient}, but this study assumes that the bandwidth allocated to all tasks for offloading is fixed. A recent survey provides a comprehensive list of studies that consider deadline-constrained problems under various settings~\cite{ramanathan2020survey}.
Thus, to the best of our knowledge, there is no study in the literature that considers the problem setting of allocating varying bandwidth and computation resource to tasks by units (access points and servers) at different locations, while having an end-to-end deadline requirement.


\section{System Model and Problem Formulation}\label{chap2-architecture}
\subsection{Edge-Cloud System Model}
\begin{figure}[tb]
	\centering
	\includegraphics[page=1, width=0.9\linewidth]{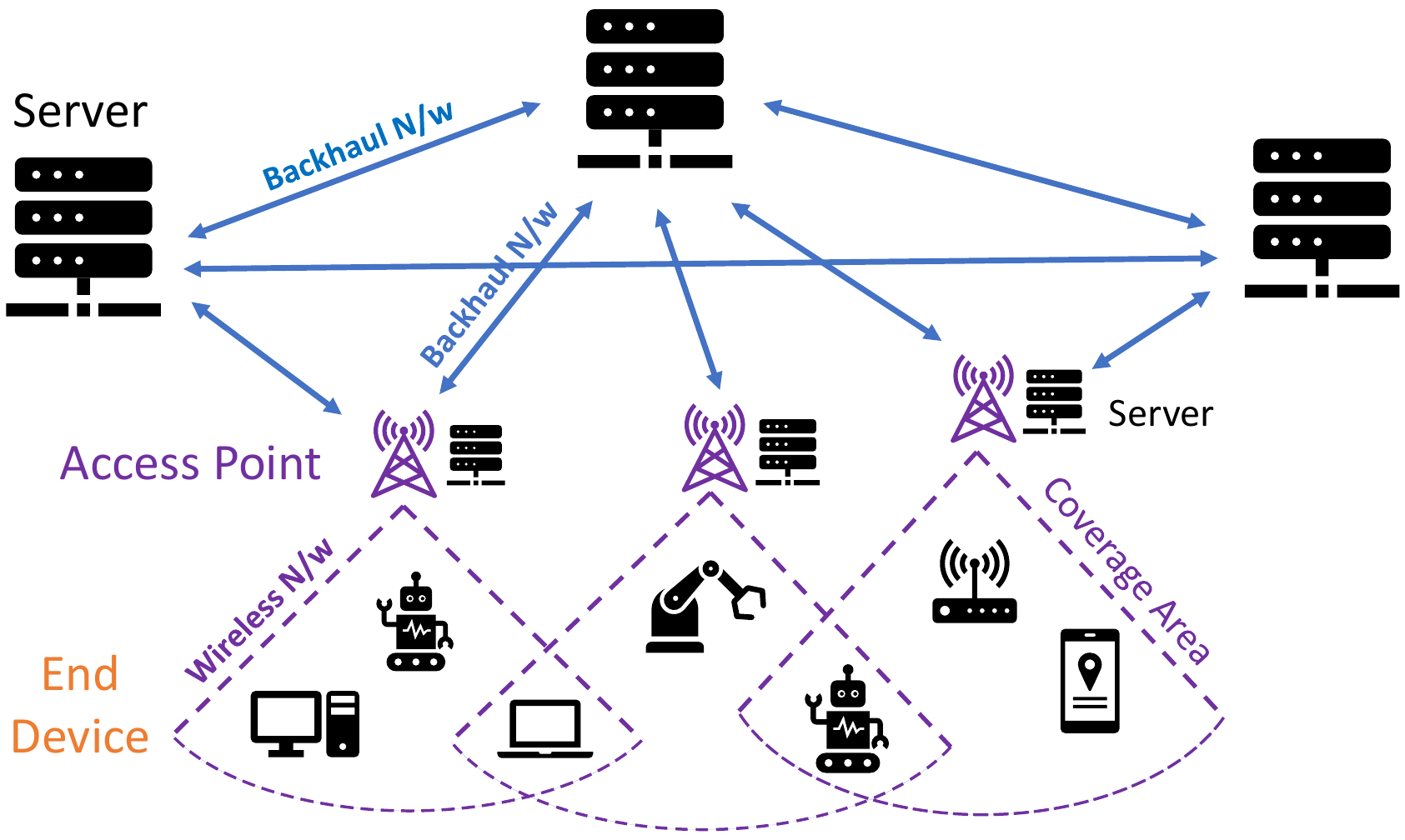}
	\caption{Edge-Cloud System Model}
	\label{fig:cloud}
\end{figure}

The multi-layer edge-cloud system comprises end devices, access points and servers, as shown in Fig.\ref{fig:cloud}. \textit{End devices} are machines that have specific functionalities and can communicate with access points through a wireless network. We denote the set of tasks generated by end devices as $\mathcal{I}$. Each task $i$, where $i \in \mathcal{I}$, has four associated parameters: $\{s_i, q_i, \Delta_i, p_i\}$. $s_i$ is the amount of data to be offloaded by task $i$, $q_i$ is the total number of CPU cycles required for task $i$, $\Delta_i$ is the task end-to-end deadline, and $p_i$ is the profit gained by completing task $i$ before its deadline. If task $i$ misses its deadline, the system will not get any profit from this task. Besides, we assume that a task cannot be split, \emph{i.e.}, it must be entirely offloaded to one access point and processed on one server. We also assume that the tasks can only be processed on servers, and therefore must necessarily be offloaded to derive profit. 

\textit{Access Points} are located near end devices 
and can collect tasks from end devices within their coverage area through wireless communication. After receiving tasks from end devices, access points will forward these tasks to servers for processing through the backhaul network. We denote the access point set as $\mathcal{A}$. Each task is covered only by a subset of access points, and we denote the access point subset that supports task $i$ offloading as $\mathcal{A}_{i}$. The wireless bandwidth capacity of an access point $j \in \mathcal{A}$ is denoted as $b_j$.

\textit{Servers} are units that process tasks forwarded from access points through a backhaul network. Compared with end devices, servers have a much greater computation resource capacity and can assist the end devices in processing computation intensive tasks.
Note that the servers that are deployed collectively with access points are called edge servers, and the servers that are deployed far away from access points are called cloud servers. Compared with cloud servers, edge servers can provide a quicker response to end devices, but usually have a smaller computation resource capacity. 
We denote the server set as $\mathcal{N}$. The computation resource capacity of a server $k \in \mathcal{N}$ is denoted as $c_k$. 

We assume that the backhaul network between access points and servers is wired, and therefore can support data transmission with constant delay irrespective of the data size, \emph{i.e.}, the available bandwidth in this network is sufficiently large. We denote the transmission delay between an access point $j$ and a server $k$ as $\delta_{jk}$, where $\delta_{jk} = \delta_{kj}$ and $\delta_{jk} = 0$ when access point $j$ and server $k$ are collectively deployed.


\subsection{Problem Formulation}\label{sec:2b}
We use a binary variable $x_{ij}$ to denote the offloading decision for task $i$. $x_{ij} = 1$ if and only if task $i$ is offloaded to access point $j \in \mathcal{A}_{i}$. Similarly, we use a binary variable $y_{ik}$ to denote the processing decision of task $i$. $y_{ik} = 1$ if and only if task $i$ is processed on server $k \in \mathcal{N}$. Furthermore, we use the variable $b_{ij}$ to denote the amount of wireless bandwidth that will be assigned to task $i$ by access point $j$, and the variable $c_{ik}$ to denote the amount of computation resource that will be assigned to task $i$ by server $k$. A summary of the notation used in this paper is provided in Table \ref{tab:notation-table}.

The total time taken to complete a task $i$, denoted as $T_i$, consists of four parts: task offloading time $T_i^o$, data transmission time in the backhaul network $T_i^c$, task processing time $T_i^p$, and the result return time. Normally, the data size of the result is negligible, so the time spent for result downloading from access points to end devices is assumed constant, and is deducted from the task deadline.
In other words, the result return time is assumed to be equal to $T^c_i$ after the result downloading time is deducted from the task deadline. If task $i$ is offloaded to  access point $j$ and processed on  server $k$, $T_i^o = {s_i}/{b_{ij}}$, $T^c_i = \delta_{jk}$, and $T^p_i = {q_i}/{c_{ik}}$. Thus, the total time taken to complete task $i$ is given as $T_i = T^o_i + 2\times T^c_i + T^p_i ={s_i}/{b_{ij}} + 2\delta_{jk} + {q_i}/{c_{ik}}$.

Thus, the deadline-constrained task mapping $\{x_{ij}, y_{ik}\}$ and resource allocation $\{b_{ij}, c_{ik}\}$ problem that aims to maximize the total system profit can be formulated as follows.
\begin{equation}\label{eq1}
	(\mathbf{P}_0) \ \ \max \sum_{i \in \mathcal{I}}\sum_{j \in \mathcal{A}_i}\sum_{k \in \mathcal{N}}x_{ij}y_{ik}p_i
\end{equation}
subject to:
\addtocounter{equation}{-1}
\begin{subequations}
	\allowdisplaybreaks
	\begin{align}
	   \begin{split}
		\sum_{j \in \mathcal{A}_{i}}x_{ij}\dfrac{s_i}{b_{ij}} + 2\sum_{j \in \mathcal{A}_{i}}\sum_{k \in \mathcal{N}}x_{ij}y_{ik}\delta_{jk} \\
		 +  \sum_{k \in \mathcal{N}}y_{ik}\frac{q_i}{c_{ik}}  \le \Delta_i ,& \ \ \forall i \in \mathcal{I} \label{eq1a}
	   \end{split}
		\\
		\sum_{j \in \mathcal{A}_{i}}x_{ij} \le 1 ,& \ \ \forall i \in \mathcal{I} \label{eq1b}
		\\
		\sum_{j \in \mathcal{A} \setminus \mathcal{A}_{i}}x_{ij} = 0 , &\ \ \forall i \in \mathcal{I} \label{eq1c}
		\\
		\sum_{k \in \mathcal{N}}y_{ik} \le 1 , &\ \ \forall i \in \mathcal{I} \label{eq1d}
		\\
		\sum_{i \in \mathcal{I}}x_{ij}b_{ij} \le b_j ,& \ \ \forall j \in \mathcal{A} \label{eq1e}
		\\
		\sum_{i \in \mathcal{I}}y_{ik}c_{ik} \le c_k ,& \ \ \forall k \in \mathcal{N} \label{eq1f}
		\\
		x_{ij} , y_{ik} \in \{0,1\}, \ \forall i \in \mathcal{I}, \ \forall j \in
		\mathcal{A},& \ \forall k \in \mathcal{N} \label{eq1g}
	\end{align}
\end{subequations}
The constraint \eqref{eq1a} guarantees that the total completion time of a task cannot exceed its deadline. Constraints \eqref{eq1b} and \eqref{eq1c} ensure that a task can only be offloaded to at most one access point in $\mathcal{A}_{i}$. Constraint \eqref{eq1d} guarantees that a task can only be processed on at most one server. Finally, constraints \eqref{eq1e} and \eqref{eq1f} ensure that the total bandwidth or computation resource assigned to all the tasks by an access point or a server cannot exceed its bandwidth or computation resource capacity.
Because of the quadratic terms $x_{ij}y_{ik}, x_{ij}b_{ij}$ and $y_{ik}c_{ik}$ in Eqs. \eqref{eq1}, \eqref{eq1a}, \eqref{eq1e} and \eqref{eq1f}, the nonconvex terms $(x_{ij}\frac{1}{b_{ij}}$ and $y_{ik}\frac{1}{c_{ik}})$ in Eq.\eqref{eq1a}, and the binary variables $x_{ij}$ and $y_{ik}$, problem $\mathbf{P}_0$ is a nonconvex MINLP optimization problem.

\begin{table}[t]
	\caption{Notation (Parameters and Variables)}
	\label{tab:notation-table}
	\begin{tabular}{|p{0.1\linewidth}|p{0.8\linewidth}|}
		\hline
		Notation & Definition \\ \hline
		$\mathcal{I}$ & Taskset, where $i \in \mathcal{I}$ denotes a task  \\ \hline
		$\mathcal{A}$ & Access point set, where $j \in \mathcal{A}$ denotes an access point \\ \hline
		$\mathcal{A}_{i}$ & Access point subset to which task $i$ can be offloaded  \\ \hline
		$\mathcal{N}$ & Server set, $k \in \mathcal{N}$ denotes a server \\ \hline
		$s_i$ &  Data size of task $i$ \\ \hline
		$q_i$  & Total number of CPU cycles needed by task $i$ \\ \hline
		$\Delta_{i}$   &  Deadline of task $i$ \\ \hline
		$p_i$    & Profit gained by completing task $i$ within deadline \\ \hline
		$b_j$ &  Bandwidth capacity of access point $j$ \\ \hline
		$c_k$    & Computation resource capacity of server $k$  \\ \hline
		$\delta_{jk}$ & Transmission delay between access point $j$ and server $k$ \\ \hline \hline
		$b_{ij}$ & Variable for wireless bandwidth assigned to task $i$ by access point $j$  \\ \hline
		$c_{ik}$  & Variable for computation resource assigned to task $i$ by server $k$  \\ \hline
		$x_{ij}$  & Binary offloading decision variable, $x_{ij} = 1$ if task $i$ is offloaded to access point $j$  \\ \hline
		$y_{ik}$  & Binary processing decision  variable, $y_{ik} = 1$ if task $i$ is processed on server $k$ \\ \hline
	\end{tabular}
\end{table}

\section{Zero-Slack based Greedy Heuristic (ZSG)}

The zero-slack based greedy heuristic algorithm (ZSG) for solving problem $\mathbf{P}_0$ comprises three main steps. First, the total available time for each task $i$ (its deadline $\Delta_i$) is distributed into three parts: task offloading time $T_i^o$, data transmission time through the backhaul network $2\times T_i^c$, and task processing time $T_i^p$. Given this distribution of total time, we calculate the required bandwidth and computation resource for each task $i$ and for every possible access point-server pair $(j,k)$. Finally, we prioritize all the $(i,j,k)$ options based on a metric, and greedily allocate them whenever feasible.
In ZSG, the resources allocated to each provisioned task are just enough for the task to be completed exactly at its deadline. 
This is possible because any feasible solution of problem $\mathbf{P}_0$ can be converted to a feasible solution that every provisioned task is completed at its deadline without any profit loss.

\textit{Deadline Distribution}:
The deadline of task $i$ is distributed into three parts: $T_i^o$, $2\times T_i^c$, and $T_i^p$.
Since the resources are allocated to tasks in the way that every provisioned task is completed exactly at its deadline, $T_i^o + T_i^p = \Delta_{i} - 2\times T_i^c$. Suppose $\gamma_{ijk}$ denotes the fraction of $T_i^o + T_i^p$ used for task $i$ offloading for a given access point-server pair $(j,k)$. That is, $T_i^o = \gamma_{ijk}(\Delta_{i} - 2\times T_i^c)$ and $T_i^p = (1 - \gamma_{ijk})(\Delta_{i} - 2\times T_i^c)$. 

In ZSG, the value of $\gamma_{ijk}$ is given by,
\begin{equation}\label{split_value_equation}
	\frac{\gamma_{ijk}}{1-\gamma_{ijk}} = \frac{(\frac{s_i}{\Delta_i})/b_j}{(\frac{q_i}{\Delta_i})/c_k}, \ \forall i \in \mathcal{I}, \ \forall j \in \mathcal{A}_{i},  \ \forall k \in \mathcal{N}
\end{equation}
The main idea for determining $\gamma_{ijk}$ is that a task with a relatively larger data size will require more time for task offloading, and a task that needs relatively more CPU cycles will require more time for task processing. 

\textit{Resource allocation calculation}:
There exist many possible access point-server pairs $(j,k)$ to offload and process a task $i$.  Due to variations in $\delta_{jk}$, the corresponding values for the required bandwidth ($b_{ij}$) and computation  resource ($c_{ik}$) to meet the task deadline are also different for different $(j,k)$ pair.
For an access point-server pair $(j,k)$, suppose $b_{ijk}$ and $c_{ijk}$ denote the bandwidth and computation resource that must be allocated to task $i$ for finishing the task at its deadline. For a given $\gamma_{ijk}$, to meet the end-to-end deadline $\Delta_i$ of task $i$, $b_{ijk}$ and $c_{ijk}$ can be calculated as follows.
\begin{equation}
	\label{bandwidth_utilization}
	b_{ijk} = \frac{s_{i}}{\gamma_{ijk}(\Delta_{i}-2\delta_{jk}) }, \ \forall i \in \mathcal{I}, \ \forall j \in \mathcal{A}_{i},  \ \forall k \in \mathcal{N}
\end{equation}

\begin{equation}
	\label{compute_utilization}
	c_{ijk} = \frac{q_{i}}{(1-\gamma_{ijk})(\Delta_{i}-2\delta_{jk}) }, \ \forall i \in \mathcal{I}, \ \forall j \in \mathcal{A}_{i},  \ \forall k \in \mathcal{N}
\end{equation}


\begin{algorithm}[h]
	\caption{Zero-Slack based Greedy Algorithm (ZSG)}\label{alg:zsg}
	\begin{algorithmic}[1]
		\REQUIRE $\mathcal{I}, \mathcal{A}, \mathcal{N}, \delta$
		\STATE $\mathcal{P} \leftarrow \emptyset$;
		\FOR{$i \in \mathcal{I}, j \in \mathcal{A}_{i}, k \in \mathcal{N}$}
			\STATE Calculate $\gamma_{ijk}, b_{ijk}, c_{ijk}$ based on Eqs. \eqref{split_value_equation}, \eqref{bandwidth_utilization}, and \eqref{compute_utilization};
			\STATE Calculate $p_{ijk}$ based on Eq. \eqref{priority}, $\mathcal{P} \leftarrow \mathcal{P} \cup \{p_{ijk}\}$;
		\ENDFOR
		\STATE Sort all $p_{ijk}$ values of set $\mathcal{P}$ in non-increasing order;
		\WHILE{$\mathcal{P} \neq \emptyset$}
		\STATE Determine the $(i, j, k)$ option with the largest $p_{ijk} \in \mathcal{P}$;
		\IF{$b_{ijk} \le b_j \ \text{and} \ c_{ijk} \le c_k$}
			\STATE $x_{ij} \leftarrow 1, y_{ik} \leftarrow 1, b_{ij} \leftarrow b_{ijk}, c_{ik} \leftarrow c_{ijk}$;
			\STATE $b_j \leftarrow b_j - b_{ijk}, c_k \leftarrow c_k - c_{ijk}$;
			\STATE $\mathcal{P} \leftarrow \mathcal{P} \setminus \{p_{i'j'k'} | i'=i, p_{i'j'k'} \in \mathcal{P}\}$;
		\ELSE
			\STATE $\mathcal{P} \leftarrow \mathcal{P} \setminus \{p_{ijk}\}$;
		\ENDIF
		\ENDWHILE
		\RETURN $\mathbf{x}, \mathbf{y}, \mathbf{b}, \mathbf{c}$
	\end{algorithmic}
\end{algorithm}

\textit{Prioritization of tasks and server pairs}:
Suppose option $(i,j,k)$ denotes the mapping of task $i$ to the access point-server pair $(j,k)$. The priority of option $(i,j,k)$ is denoted as $p_{ijk}$, and given by the following equation.
\begin{equation}
	\label{priority}
	p_{ijk} = \frac{p_{i}}{ \left ( \frac{b_{ijk}}{b_{j}} \right ) \times \left ( \frac{c_{ijk}}{c_{k}} \right ) }, \ \forall i \in \mathcal{I}, \ \forall j \in \mathcal{A}_{i},  \ \forall k \in \mathcal{N}
\end{equation}
The intuition behind this metric is that an option $(i,j,k)$ with a higher profit $(p_i)$ and lower resource usage $(b_{ijk} \text{ and } c_{ijk})$ should be given higher priority. 

The detail steps of ZSG are presented in Algorithm \ref{alg:zsg}. For each $(i,j,k)$ option, calculate $\gamma_{ijk}, b_{ijk}$ and $ c_{ijk}$ based on Eqs. \eqref{split_value_equation}, \eqref{bandwidth_utilization} and \eqref{compute_utilization}, where $i \in \mathcal{I}, j \in \mathcal{A}_i, k \in \mathcal{N}$ (line 3). Then, calculate $p_{ijk}$ according to Eq. \eqref{priority}, and add $p_{ijk}$ to set $\mathcal{P}$ (line 4). After all possible $p_{ijk}$ values are calculated, sort these $p_{ijk}$ values in nonincreasing order (line 6). If set $\mathcal{P}$ is not empty, the $(i,j,k)$ option with the largest priority value $(p_{ijk})$ is chosen, and the corresponding mapping and allocation are realized if the resource capacity constraints on access point $j$ and server $k$ are met (lines 7-11). Once a  task is provisioned, all the $p_{ijk}$ values related to task $i$ are removed from set $\mathcal{P}$ (line 12). Otherwise, only $p_{ijk}$ is discarded from set $\mathcal{P}$ and the algorithm proceeds with the next $(i,j,k)$ option with largest $p_{ijk}$ value in set $\mathcal{P}$ (line 14). The algorithm stops when set $\mathcal{P}$ is empty.




\section{Linear Discretization Method (LDM)}\label{chap3-methodology}
In this section, we present a linear discretization method (LDM) to solve the nonconvex MINLP problem $\mathbf{P}_0$ by reformulating it to an ILP problem.
The LDM assumes that minimum units exist for the allocation of bandwidth and computation resource and any resource allocation will be an integer multiple of corresponding minimum unit. Suppose the minimum unit of bandwidth is denoted as $\tilde{b}$ and that for the computation resource is denoted as $\tilde{c}$. In Problem $\mathbf{P}_0$, we now replace terms as follows: $b_{ij}$ by $u_{ij}\tilde{b}$ and $c_{ik}$ by $v_{ik}\tilde{c}$ where $u_{ij}$ and $v_{ik}$ are nonnegative integer variables, and $b_j$ by $u_j\tilde{b}$ and $c_k$ by $v_k\tilde{c}$ where $u_j$ and $v_k$ are positive integer parameters. Note that $u_j$ and $v_k$ are the upper bounds for $u_{ij}$ and $v_{ik}$, respectively, for all $i \in \mathcal{I}$.

In the discretized version of problem $\mathbf{P}_0$ (as defined above), the deadline constraint of Eq.\eqref{eq1a} can be rewrite as follows.
\begin{equation}\label{eq:idm0}
    \begin{split}
		\sum_{j \in \mathcal{A}_{i}}x_{ij}\dfrac{s_i}{u_{ij}\tilde{b}} + 2\sum_{j \in \mathcal{A}_{i}}\sum_{k \in \mathcal{N}}x_{ij}y_{ik}\delta_{jk} \\
		 +  \sum_{k \in \mathcal{N}}y_{ik}\frac{q_i}{v_{ik}\tilde{c}}  \le \Delta_i ,& \ \ \forall i \in \mathcal{I}
	   \end{split}
\end{equation}
Eq.~\eqref{eq:idm0} is still nonconvex because of the terms $x_{ij}\frac{1}{u_{ij}}$ and $y_{ik}\frac{1}{v_{ik}}$. To linearize these terms, the general idea is to discretize one variable and use the summation of finite linear terms to replace the original nonconvex term \cite{koster2019adaptive}. Take $x_{ij}\frac{1}{u_{ij}}$ as an example.
The variable ${u_{ij}}$ is mapped into a finite number of possible values; this is feasible because $u_{ij}$ is an integer with a finite range. Each positive $u_{ij}$ value is associated with a new binary variable $x_{ijm} \in \{0,1\}$, $m \in \{ 1,2, \ldots, u_j \}$, where $\sum_{m=1}^{u_j}x_{ijm} = x_{ij} \leq 1$. The variable $x_{ijm}$ determines which discrete value $m$ is chosen by $u_{ij}$. $x_{ijm} = 1$ only when the discrete value $m$ is selected and in this case $x_{ij}\frac{1}{u_{ij}} = x_{ijm}\frac{1}{m}$. For the case when $x_{ij} = 0$, $\sum_{m=1}^{u_j}x_{ijm} = 0$ and none of the discrete values is selected. Note, when $u_{ij}=0$, $x_{ij}$ must be $0$ to satisfy the deadline constraint~\eqref{eq:idm0} and in this case as well $\sum_{m=1}^{u_j}x_{ijm} = 0$.
Thus, the nonconvex term $x_{ij}\frac{1}{u_{ij}}$ can be redefined as follows.
\begin{equation}\label{eq:idm1}
    x_{ij}\frac{1}{u_{ij}} = \sum_{m=1}^{u_j}x_{ijm}\frac{1}{m},  \ \forall i \in \mathcal{I}, \ \forall j \in \mathcal{A}_{i}
\end{equation}


After the discretization of $u_{ij}$, we have
\begin{equation}\label{eq:idm2}
	u_{ij} = \sum_{m=1}^{u_j}x_{ijm}m \le \sum_{m=1}^{u_j}x_{ijm}u_j \le x_{ij}u_j.
\end{equation}

Using the same technique, we can also linearize the term $y_{ik}\frac{1}{v_{ik}}$. Suppose each positive value of ${v_{ik}}$ is associated with a new binary variable $y_{ikn} \in \{0,1\}$, $n \in \{ 1,2, \ldots, v_k \}$, where $\sum_{n=1}^{v_k}y_{ikn} = y_{ik} \leq 1$. Then, the nonconvex term $y_{ik}\frac{1}{v_{ik}}$ can be redefined as follows.
\begin{equation}\label{eq:idm3}
	y_{ik}\frac{1}{v_{ik}} = \sum_{n=1}^{v_k}y_{ikn}\frac{1}{n} ,  \ \forall i \in \mathcal{I}, \ \forall k \in \mathcal{N}.
\end{equation}
\begin{equation}\label{eq:idm4}
	v_{ik} = \sum_{n=1}^{v_k}y_{ikn}n \le \sum_{n=1}^{v_k}y_{ikn}v_k \le y_{ik}v_k.
\end{equation}

Eqs. \eqref{eq:idm2} and \eqref{eq:idm4} define the property that when task $i$ is not mapped to access point $j$ or server $k$, where $x_{ij} = 0$ or $y_{ik} = 0$, no corresponding resource will be assigned to task $i$, which gives $u_{ij} = 0$ or $v_{ik} = 0$.
Thus, constraints \eqref{eq1e} and \eqref{eq1f} in the discretized version of Problem $\mathbf{P}_0$ can be rewritten as follows.
\begin{equation}\label{eq:idm5a}
    \sum_{i \in \mathcal{I}}x_{ij}u_{ij} = \sum_{i \in \mathcal{I}}u_{ij} = \sum_{i \in \mathcal{I}}\sum_{m=1}^{u_j}x_{ijm}{m} \le u_j,\ \forall j \in \mathcal{A}
\end{equation}
\begin{equation}\label{eq:idm5b}
    \sum_{i \in \mathcal{I}}y_{ik}v_{ik} = \sum_{i \in \mathcal{I}}v_{ik} = \sum_{i \in \mathcal{I}}\sum_{n=1}^{v_k}y_{ikn}n \le v_k,\ \forall k \in \mathcal{N}
\end{equation}

Thus far, we have transformed the nonconvex terms $x_{ij}\frac{1}{u_{ij}}$ and $y_{ik}\frac{1}{v_{ik}}$ into linear terms. In problem $\mathbf{P}_0$, there still exists the quadratic term $x_{ij}y_{ik}$ in the objective function \eqref{eq1} as well as in constraint \eqref{eq:idm0}.
For the quadratic term $x_{ij}y_{ik}$, we use a new binary variable $z_{ijk} \in \{0,1\}$ to replace it. $z_{ijk} = 1$ only when task $i$ is offloaded to access point $j$ ($x_{ij} = 1$) and processed on server $k$ ($y_{ik} = 1$). Since $\sum_{m=1}^{u_j}x_{ijm} = x_{ij}$ and $\sum_{n =1}^{v_k}y_{ikn} = y_{ik}$, the binary variable $z_{ijk}$ can be defined using the following linear constraints \cite{minlp2016}.
\begin{equation}\label{eq:idm6b}
    z_{ijk} \ge \sum_{m=1}^{u_j}x_{ijm} + \sum_{n=1}^{v_k}y_{ikn} - 1 ,   \forall i \in \mathcal{I},  \forall j \in \mathcal{A}_{i},  \forall k \in  \mathcal{N}
\end{equation}
\begin{equation}\label{eq:idm6c}
    z_{ijk} \le \sum_{m=1}^{u_j}x_{ijm} ,   \forall i \in \mathcal{I},  \forall j \in \mathcal{A}_{i}, \forall k \in  \mathcal{N}
\end{equation}
\begin{equation}\label{eq:idm6d}
    z_{ijk} \le \sum_{n=1}^{v_k}y_{ikn} ,   \forall i \in \mathcal{I},  \forall j \in \mathcal{A}_{i},  \forall k \in  \mathcal{N}
\end{equation}
\begin{equation}\label{eq:idm6a}
    z_{ijk} \in \{0,1\},   \forall i \in \mathcal{I},  \forall j \in \mathcal{A}_{i},  \forall k \in  \mathcal{N}
\end{equation}

Eq.\eqref{eq:idm6b} ensures that $z_{ijk}$ is $1$ only when both $x_{ij}$ and $y_{ik}$ are $1$. Thus, the nonconvex MINLP problem $\mathbf{P}_0$ can be reformulated as an ILP problem, under the assumption that resources are allocated in integer multiples of minimum resource units.
Note that in the ILP problem, for given resource capacities ($b_j$ and $c_k$ values), smaller values for $\tilde{b}$ and $\tilde{c}$ will result in increased values for $u_j$ and $v_k$, thus increasing the number of variables. Although this can improve solution quality, it will also lead to increased runtime.

\section{Experiment}\label{chap4-experiment}
In this section, we present the experimental results that evaluate the performances of ZSG and LDM. We generate a variety of synthetic tasksets with different parameter settings and provision them on a fixed edge-cloud architecture. The algorithms are compared in terms of achieved system profit.

\begin{table}[tb]
	\caption{Ranges used for Various Parameters}
	\label{tab:Ranges for Sampling}
	\centering
	\begin{tabular}{|p{0.3\linewidth}|p{0.45\linewidth}|}
		\hline
		Parameter & Range \\ \hline
        $c_k, k \in \text{Cloud Servers} $ & $80$ to $100$ \\ \hline
        $c_k, k \in \text{Edge Servers}$ & $40$ to $60$ \\ \hline
        $b_j, j \in \mathcal{A}$ & $40$ to $100$ \\ \hline
        $\delta_{jk}, j \in \mathcal{A}, k \in \mathcal{N}$ &  $0$ to $10$ \\ \hline
        $|\mathcal{A}_{i}|, i \in \mathcal{I}$ & $1$ to $2$ \\ \hline
        $p_i, i \in \mathcal{I}$ & $10$ to $100$ \\ \hline
        $\Delta_i, i \in \mathcal{I}$ & $2\times \delta_{jk}^{\max}+15$ to $2\times \delta_{jk}^{\max}+45$, or $2\times \delta_{jk}^{{mean}}+15$ to $2\times \delta_{jk}^{{mean}}+45$\\ \hline
	\end{tabular}
\end{table}

\subsection{Taskset Generation}
The number of access points and servers in the system are fixed at $20$ and $25$, respectively. $20$ of the $25$ servers are edge servers, which are deployed collectively with access points, and the remaining $5$ servers are cloud servers.
Additionally, all the parameters related to access points and servers, including $c_k, b_j$ and $\delta_{jk}$, are randomly sampled integer values from a pre-defined range in Table~\ref{tab:Ranges for Sampling}. Note that only the $\delta_{jk}$ between the collectively deployed access point $j$ and server $k$ is set to $0$. Cloud servers have larger computation resource capacity than that of edge servers. Thus, the capacity range of the cloud servers is from $80$ to $100$, and the capacity range of the edge servers is from $40$ to $60$. 
These values are then kept fixed throughout the experiments\footnote{Although the edge-cloud architecture parameters are fixed in all our experiments, the taskset parameters are varied across a wide range to evaluate the performance of the algorithms for different resource usage scenarios.}.

For each task $i \in \mathcal{I}$, its profit $p_i$, deadline $\Delta_i$, and the number of access points supporting task $i$ offloading ($|\mathcal{A}_{i}|$) are also randomly sampled integer values from a pre-defined range as shown in Table~\ref{tab:Ranges for Sampling}. 
For cost concern, the deployment of access points should avoid too many overlapping coverage areas, thus, the number of access points each task can be offloaded to ranges from $1$ to $2$.
Given $|\mathcal{A}_{i}|$, the access points to which a task can offload are randomly chosen from the $20$ access points. Suppose $\delta_{jk}^{\max} = \max_{j \in \mathcal{A}, k \in \mathcal{N}}(\delta_{jk})$ and $\delta_{jk}^{{mean}} = {mean}_{j \in \mathcal{A}, k \in \mathcal{N}}(\delta_{jk})$. The task deadlines are sampled from two different ranges with equal probability. 
One of them, $[2\times \delta_{jk}^{\max}+15, 2\times \delta_{jk}^{\max}+45]$, uses the largest access point to server transmission delay, representing tasks that have relatively more time for offloading and processing. Whereas the other, $[2\times \delta_{jk}^{{mean}} +15, 2\times \delta_{jk}^{{mean}}+45]$, uses the average access point to server transmission delay, and thus represents tasks that have relatively less time for offloading and processing. These task parameters are sampled repeatedly when generating the tasks in each taskset.

To synthesize tasksets with varying levels of resource usage, we generate tasks with varying bandwidth and computation resource utilizations (amount of resource required in a given time interval). This in effect varies the $q_i$ and $s_i$ values for each task $i$. For the time interval, we use $\tau_i=\Delta_i - 2\times \delta_{jk}^{{mean}}$, which roughly captures the amount of time available to complete both offloading and processing. Thus, for wireless bandwidth, the utilization of a task $i$ offloading to an access point $j$, $ub_{ji}$, is defined as $ub_{ji} = s_{i}/(b_j \times \tau_i)$. Similarly, for computation resource, the utilization of a task $i$, $uc_{i}$, is defined as $uc_{i} = q_{i}/(\min_{k \in \mathcal{N}} c_k\times \tau_i)$. For feasibility, we assume $ub_{ji}$ and $uc_i$ are always less than or equal to $1$.  

To generate tasksets, we consider a different number of tasks in each taskset ($\{ 40, 60, 80, 100, 120 \}$). For the wireless bandwidth, we consider different values for the total bandwidth utilization of each access point ($ub \in \{ 0.3, 0.4, 0.5, 0.6, 0.7, 0.8, 0.9, 1 \}$).
Similarly, for the computation resource, we consider different values for the total compute utilization of the entire edge-cloud system ($uc \in \{ 1, 5, 9, 13, 17, 21, 25, 29, 33, 35.50\}$), where $35.50 = \sum_{k \in \mathcal{N}} c_k / \min_{k \in \mathcal{N}} c_k$ since we normalize $uc$ by $\min_{k \in \mathcal{N}} c_k$.
For each combination of these three parameters ($400$ in all), we generate $30$ tasksets, resulting in a total of $12,000$ different tasksets.

\begin{figure}[b]
    \includegraphics[width=\linewidth]{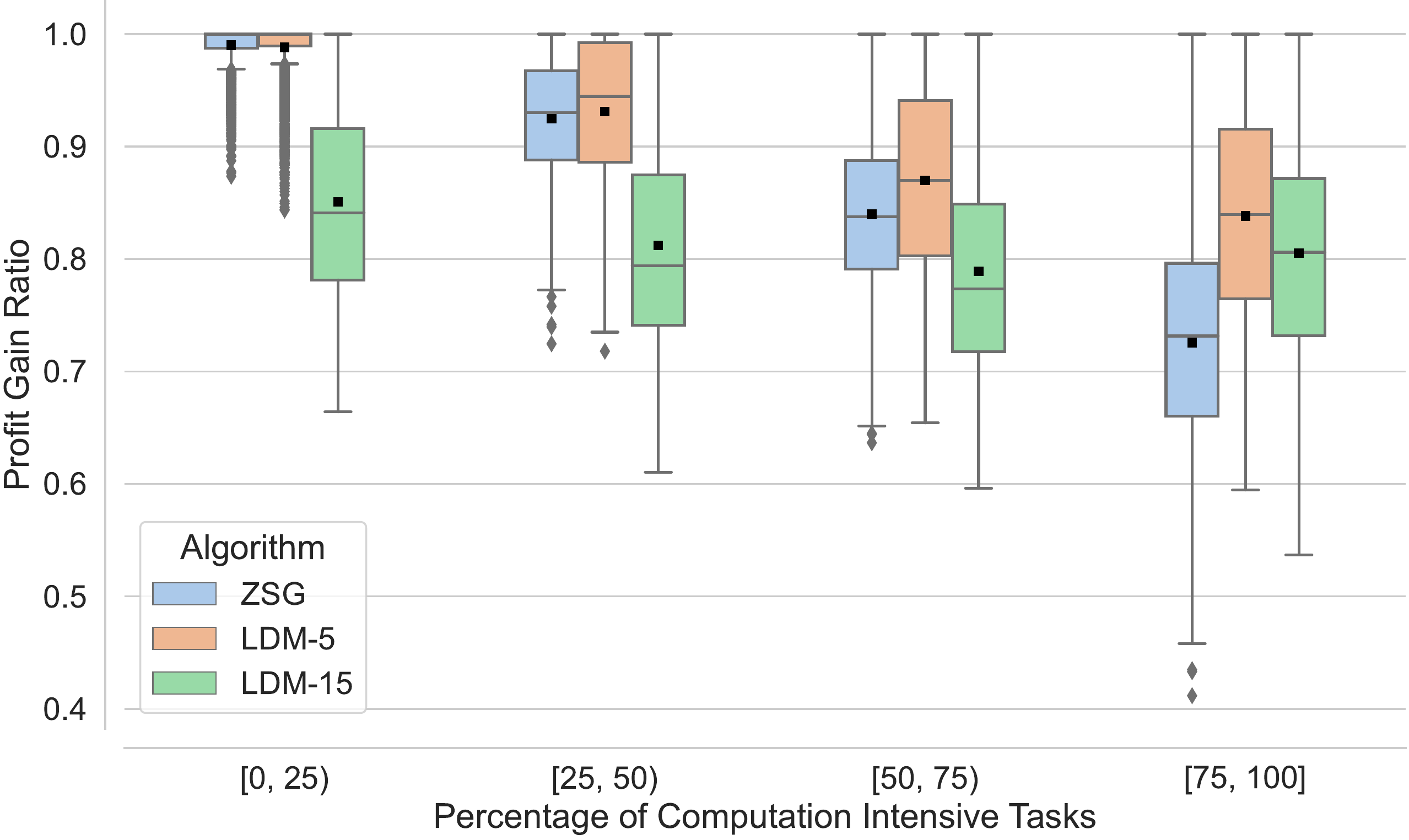}
    \caption{Performance with varying percentage of computation intensive tasks}
    \label{fig:ci}
\end{figure}

To generate a single taskset, given the values for $ub$, $uc$ and the number of tasks, we first generate the profit $p_i$, deadline $\Delta_i$ and access point set $\mathcal{A}_{i}$ for each task $i$ as described earlier. Then, for each access point $j \in \mathcal{A}$, given total utilization $ub\times b_j$ and the tasks that can be offloaded to that access point (denoted by set $\mathcal{I}_j$), we use an existing algorithm called Uunifast~\cite{bini2005measuring} to generate the task bandwidth utilization values $ub_{ji}$ such that $ub = \sum_{i \in \mathcal{I}_j} ub_{ji}$. This algorithm efficiently generates the task utilization values using uniform random sampling and without any bias. Since a task $i$ can be within the coverage area of more than one access point, it can have different $ub_{ji}$ values assigned to it by this algorithm for each feasible access point $j$. Therefore, we set its data size $s_i$ as the maximum obtained from those values given by $\max_{j \in \mathcal{A}_{i}} (ub_{ji}\times b_j\times \tau_i)$. Finally, given total computation resource utilization $uc$, we use another existing algorithm called Stafford's Randfixedsum~\cite{emberson2010techniques} to generate the tasks' computation resource utilization values $uc_i$ such that $uc = \sum_{i \in \mathcal{I}} uc_i$. This algorithm uses similar techniques as Uunifast and generates uniformly random and unbiased task utilization values even when the total computation resource utilization $uc$ is greater than $1$. We also restrict each $uc_i$ to be no more than $1$ to ensure that the compute requirement $q_i$ of every task $i \in \mathcal{I}$ can be satisfied by any server.

For LDM, we consider two different values for the minimum units ($\tilde{c}$ and $\tilde{b}$), $5$ and $15$, and denote the corresponding LDM as LDM-5 and LDM-15. Once $\tilde{c}$ and $\tilde{b}$ are fixed, $v_k$ and $u_j$ in Eqs. \eqref{eq:idm5a} and \eqref{eq:idm5b}, are set as $\lfloor{c_k/\tilde{c}}\rfloor$ and $\lfloor{b_j/\tilde{b}}\rfloor$,  respectively.
Besides, since the runtimes of the algorithms are generally proportional to the number of tasks in a taskset, we also partition the tasksets based on this number and allocate runtimes to LDM-5 (likewise LDM-15) to be $600$ (likewise $200$)  times the maximum observed runtime for ZSG within each partition.
This ensures a fair comparison because LDM-5 has three times more variables than LDM-15 and the ILP solver generally requires orders of magnitude more time than the ZSG heuristic.
Experiments were run on a desktop PC with Intel Xeon(R) Gold 5220R 2.2GHz CPU and 128GB of RAM, and Gurobi was used as the ILP solver of LDM\footnote{Experiments code is available at https://github.com/CPS-research-group/CPS-NTU-Public/tree/GLOBECOM2022.}.

\subsection{Discussion of Results}
To visualize the results, we have categorized tasksets based on the resource usage intensity of tasks. A task $i$ is identified to be \emph{computation intensive} if $q_{i}/\tau_i$ is more than $20\%$ of the minimum $c_k$ value. Likewise, a task is identified to be \emph{bandwidth intensive} if $s_{i}/\tau_i$ is more than $20\%$ of the minimum $b_j$ value among all $j \in \mathcal{A}_{i}$. The \emph{profit gain ratio} is used to compare the performances of different algorithms, which is the ratio of the total profit of tasks provisioned by the algorithm to the total profit of all tasks in the taskset.
\[\text{profit gain ratio} = \frac{\text{total profit of provisioned tasks}}{\text{total profit of all tasks in the taskset}}\]

\begin{figure}[tb]
    \includegraphics[width=\linewidth]{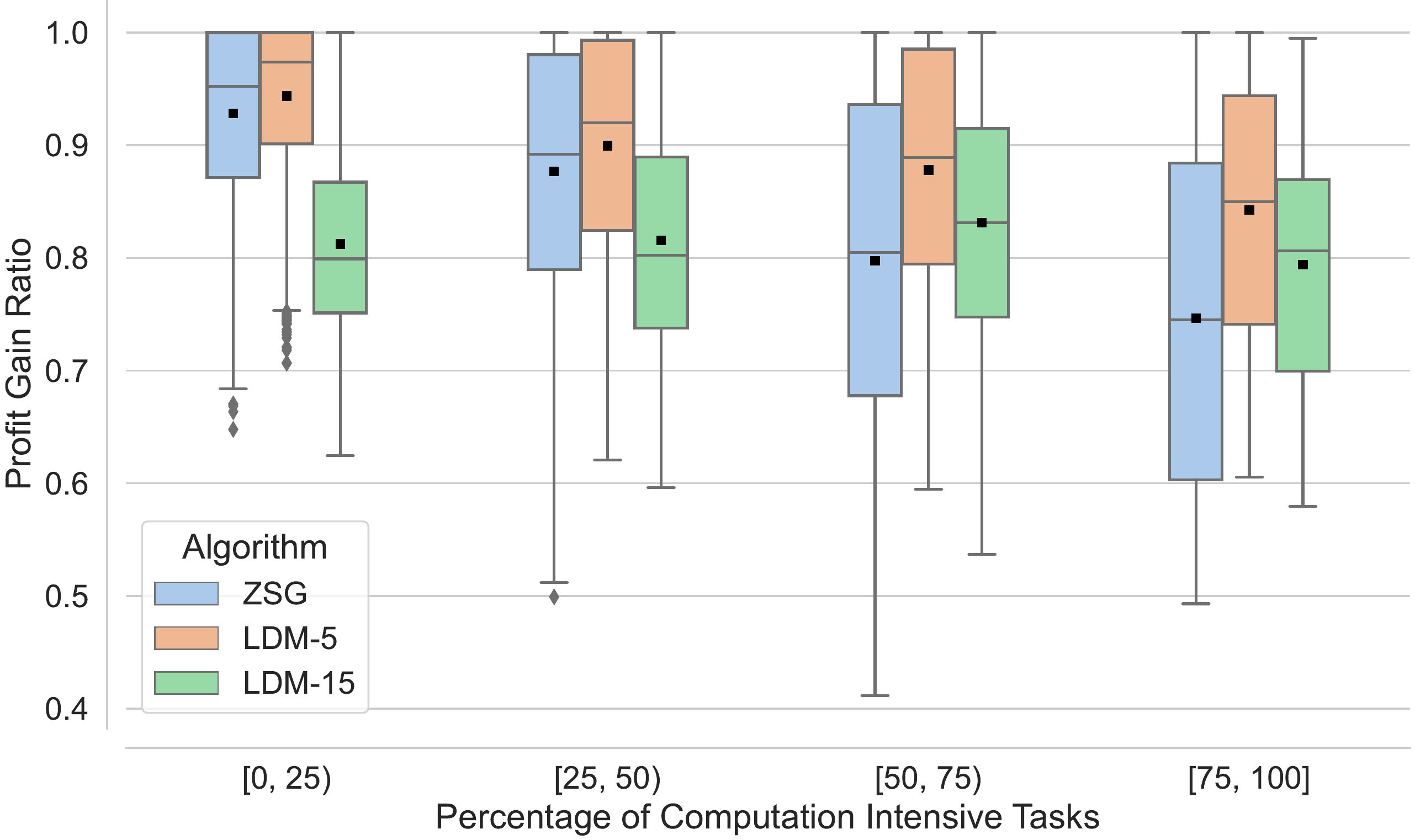}
    \caption{Performance with varying percentage of bandwidth intensive tasks}
    \label{fig:bi}
\end{figure}

\begin{figure}[tb]
    \centering
    \includegraphics[width=\linewidth]{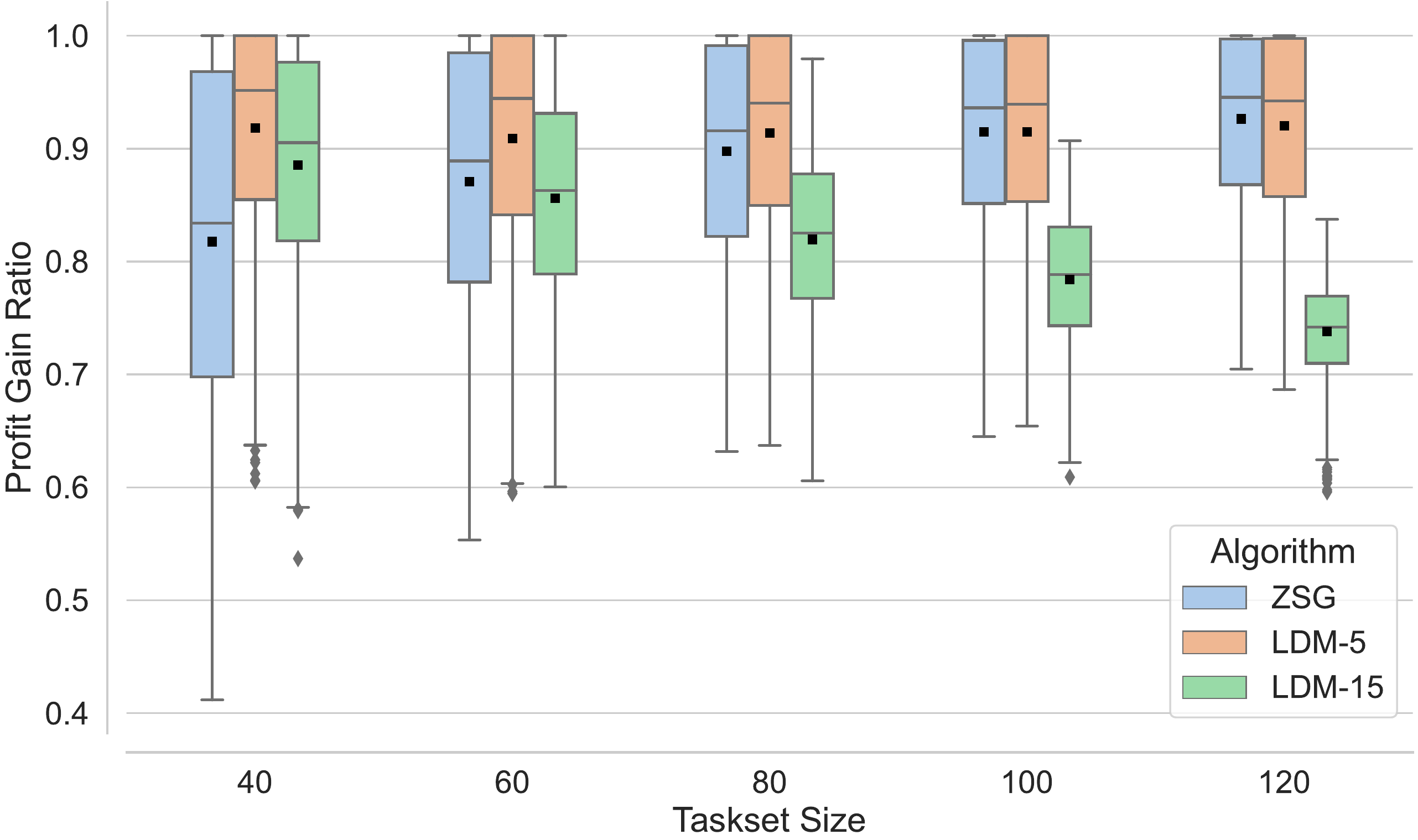}
    \caption{Performance with varying taskset size }
    \label{fig:is}
\end{figure}

Results for tasksets with varying levels of computation and bandwidth intensive tasks are shown in Figs. \ref{fig:ci} and \ref{fig:bi}, respectively. In these plots, x-axis denotes the percentage of resource intensive tasks in the taskset, and y-axis denotes the {profit gain ratio}. For each algorithm the figures show a standard box-plot, with the line (likewise dot) inside the box denoting median (likewise mean).
As can be seen from these figures, the performance of ZSG is close to LDM-5 when the percentage of resource intensive tasks is small, but this gap widens as the percentage increases. LDM-5 obtains an average of $2.98\%$ more profit than ZSG for the considered tasksets. This is expected because tasksets with a higher percentage of resource intensive tasks are usually harder to provision, and in these cases, LDM-5 with a standard ILP solver performs better.
Note that although LDM-5 outperforms ZSG, their performance gap is small. This is because the minimum resource units and running time of the ILP solver limit the performance of LDM-5.
It can also be observed that the performances of both LDM-5 and ZSG are better than LDM-15. On average, LDM-5 and ZSG obtain $9.87\%$ and $6.88\%$ more system profit than LDM-15, respectively. This shows that the performance of LDM critically depends on the granularity of discretization.


We have also compared the performance of ZSG and LDM with varying taskset sizes, and this result is shown in Fig. \ref{fig:is}. Interestingly, the performance of LDM-15 decreases significantly with the increase in taskset size, whereas that of ZSG and LDM-5 reamins the same. As the taskset size increases, the resource requirement of each task generally decreases, thus increasing the resource loss incurred in LDM due to discretization (and hence decreasing achieved profit). The larger base unit in LDM-15 results in a more significant resource loss during task provision and causes LDM-15 to perform worse than LDM-5 and ZSG.

\section{Conclusion}\label{chap5-conclusion}
This paper addressed a deadline-constrained multi-resource task mapping and allocation problem for an edge-cloud system. A key challenge of the problem was the allocation of resources on two different units (access point and server) for the same task with an end-to-end deadline. Two effective methods, called ZSG and LDM, were proposed to determine the task mapping and allocation of wireless bandwidth and computation resources. Experimental results demonstrated the efficiency of the proposed methods when dealing with tasksets with a variety of resource requirements. Although LDM with smaller minimum resource units outperformed ZSG, the ILP solver of LDM generally required orders of magnitude more time than ZSG. Thus, ZSG is superior to LDM in large-scale system, and LDM is preferred if the system has hardware acceleration for the ILP solver or requires the solution with a high profit gain ratio.

In this paper, we assume that the data transmission latency is independent of the data size in the backhaul network, which might limit the application of the proposed edge-cloud system in real-world systems.
In the future, we would like to explore an edge-cloud system where the data size affects data transmission latency in the backhaul network.
Besides, we would also like to explore a distributed and online solution for the presented problem, specifically considering resource scheduling over time.



\bibliographystyle{IEEEtran} 
\bibliography{IEEEabrv,references}

\begin{thebibliography}{10}
\providecommand{\url}[1]{#1}
\csname url@samestyle\endcsname
\providecommand{\newblock}{\relax}
\providecommand{\bibinfo}[2]{#2}
\providecommand{\BIBentrySTDinterwordspacing}{\spaceskip=0pt\relax}
\providecommand{\BIBentryALTinterwordstretchfactor}{4}
\providecommand{\BIBentryALTinterwordspacing}{\spaceskip=\fontdimen2\font plus
\BIBentryALTinterwordstretchfactor\fontdimen3\font minus
  \fontdimen4\font\relax}
\providecommand{\BIBforeignlanguage}[2]{{%
\expandafter\ifx\csname l@#1\endcsname\relax
\typeout{** WARNING: IEEEtran.bst: No hyphenation pattern has been}%
\typeout{** loaded for the language `#1'. Using the pattern for}%
\typeout{** the default language instead.}%
\else
\language=\csname l@#1\endcsname
\fi
#2}}
\providecommand{\BIBdecl}{\relax}
\BIBdecl

\bibitem{li20185g}
Z.~Li, M.~A. Uusitalo, H.~Shariatmadari, and B.~Singh, ``5g urllc: Design
  challenges and system concepts,'' in \emph{2018 15th International Symposium
  on Wireless Communication Systems (ISWCS)}, 2018, pp. 1--6.

\bibitem{chekuri2005polynomial}
C.~Chekuri and S.~Khanna, ``A polynomial time approximation scheme for the
  multiple knapsack problem,'' \emph{SIAM Journal on Computing}, vol.~35,
  no.~3, pp. 713--728, 2005.

\bibitem{vu2021optimal}
T.~T. Vu, D.~N. Nguyen, D.~T. Hoang, E.~Dutkiewicz, and T.~V. Nguyen, ``Optimal
  energy efficiency with delay constraints for multi-layer cooperative fog
  computing networks,'' \emph{IEEE Transactions on Communications}, vol.~69,
  no.~6, pp. 3911--3929, 2021.

\bibitem{li2019cooperative}
Q.~Li, J.~Zhao, and Y.~Gong, ``Cooperative computation offloading and resource
  allocation for mobile edge computing,'' in \emph{2019 IEEE International
  Conference on Communications Workshops (ICC Workshops)}, 2019, pp. 1--6.

\bibitem{millnert2018achieving}
V.~Millnert, J.~Eker, and E.~Bini, ``Achieving predictable and low end-to-end
  latency for a network of smart services,'' in \emph{2018 IEEE Global
  Communications Conference (GLOBECOM)}, 2018, pp. 1--7.

\bibitem{cziva2018dynamic}
R.~Cziva, C.~Anagnostopoulos, and D.~P. Pezaros, ``Dynamic, latency-optimal vnf
  placement at the network edge,'' in \emph{IEEE INFOCOM 2018 - IEEE Conference
  on Computer Communications}, 2018, pp. 693--701.

\bibitem{yang2019efficient}
C.~Yang, Y.~Liu, X.~Chen, W.~Zhong, and S.~Xie, ``Efficient mobility-aware task
  offloading for vehicular edge computing networks,'' \emph{IEEE Access},
  vol.~7, pp. 26\,652--26\,664, 2019.

\bibitem{ramanathan2020survey}
S.~Ramanathan, N.~Shivaraman, S.~Suryasekaran, A.~Easwaran, E.~Borde, and
  S.~Steinhorst, ``A survey on time-sensitive resource allocation in the cloud
  continuum,'' \emph{it-Information Technology}, vol.~62, no. 5-6, pp.
  241--255, 2020.

\bibitem{koster2019adaptive}
A.~M. Koster and S.~Kuhnke, ``An adaptive discretization algorithm for the
  design of water usage and treatment networks,'' \emph{Optimization and
  Engineering}, vol.~20, no.~2, pp. 497--542, 2019.

\bibitem{minlp2016}
\BIBentryALTinterwordspacing
An introduction to mixed integer nonlinear optimization. [Online]. Available:
  \url{https://www.ima.umn.edu/2015-2016/ND8.1-12.16/25419}
\BIBentrySTDinterwordspacing

\bibitem{bini2005measuring}
E.~Bini and G.~C. Buttazzo, ``Measuring the performance of schedulability
  tests,'' \emph{Real-Time Systems}, vol.~30, no.~1, pp. 129--154, 2005.

\bibitem{emberson2010techniques}
P.~Emberson, R.~Stafford, and R.~I. Davis, ``Techniques for the synthesis of
  multiprocessor tasksets,'' in \emph{proceedings 1st International Workshop on
  Analysis Tools and Methodologies for Embedded and Real-time Systems (WATERS
  2010)}, 2010, pp. 6--11.

\end{thebibliography}

\end{document}